\documentclass[a4paper,fleqn,usenatbib]{mnras}
\usepackage{newtxtext,newtxmath}
\usepackage{ae,aecompl}
\usepackage{graphicx}
\usepackage{amsmath}
\usepackage{amssymb}
\usepackage{longtable}
\usepackage[T1]{fontenc}
\usepackage{geometry}
\usepackage[labelfont=bf]{caption}
\usepackage{floatrow}
\usepackage{hyperref,natbib}
\title[High-speed photometry of faint CVs - IX.]{High-speed photometry of faint cataclysmic variables - IX. Targets from multiple transient surveys.}

\author[K. Paterson et al.]{K. Paterson,$^{1}$\thanks{E-mail: ptrker004@myuct.ac.za} P. A. Woudt,$^{1}$ B. Warner,$^{1}$  H. Breytenbach,$^{1,2}$ C. K. Gilligan,$^{3}$ 
\and M. Motsoaledi,$^{1,2}$ J. R. Thorstensen,$^{3}$ H. L. Worters$^{2}$
\\
$^{1}$Department of Astronomy, University of Cape Town, Private Bag X3, 7701, Rondebosch, South Africa\\
$^{2}$South African Astronomical Observatory, PO Box 9, Observatory 7935, Cape Town, South Africa\\
$^{3}$Department of Physics and Astronomy, Dartmouth College, Hanover NH 03755, USA}

\date{Accepted 2019 March 25. Received 2019 March 23; in original form 2018 July 19}

\pubyear{2019}

\begin{document}
\label{firstpage}
\pagerange{\pageref{firstpage}--\pageref{lastpage}}
\maketitle

\begin{abstract}
We present high-speed photometric observations of 25 cataclysmic variables detected by the All Sky Automated Search for Super-Novae \textcolor{black}{(ASAS-SN)}, the Mobile Astronomical System of the TElescope-Robot \textcolor{black}{(MASTER)} and the Catalina Real-Time Transient Survey \textcolor{black}{(CRTS)}. From these observations we determine 16 new orbital periods and 1 new superhump period. \textcolor{black}{Two} systems (ASASSN-14ik and ASASSN-14ka) have outburst periods of approximately 1 month, with a third (ASASSN-14hv) having outbursts approximately every 2 months. Included in the sample are 11 eclipsing systems, one probable intermediate polar (ASASSN-15fm), 1 SW Sex-type star (MLS 0720+17), 1 WZ Sge-type star (ASASSN-17fz) and one system showing different photometric and spectroscopic periods (ASASSN-15kw).
\end{abstract}


\section{Introduction}
We present the latest results of photometric follow-up of faint cataclysmic variables (CVs; see \citealt{1995CAS....28.....W} for a review on CVs) that are accessible from the southern hemisphere. This work is the last in series of papers (see \citealt{2014MNRAS.437..510C} and references therein) aimed at the characterisation of newly discovered CVs, including the determination of their orbital periods, a search for sub-orbital periodicities and the discovery of interesting targets for possible in-depth studies. Previous papers in the series focused on faint nova remnants and CVs identified by the Sloan Digital Sky Survey (SDSS; see \citealt{2002AJ....123..430S}, \citealt{2003AJ....126.1499S}), attention then shifted to CVs discovered by the Catalina Real-Time Transient Survey (CRTS; see \citealt{2009ApJ...696..870D}).\\
In this paper we present observations of 25 faint CVs identified in the All Sky Automated Search for Super-Novae (ASAS-SN; see \citealt{2014AAS...22323603S}), the Mobile Astronomical System of the TElescope-Robots (MASTER) node situated in Sutherland (MASTER-SAAO; see \citealt{2010AdAst2010E..30L}), as well as from CRTS. The ASAS-SN survey is a dedicated all-sky survey focusing on the search for supernovae. It is made up of \textcolor{black}{five} units, each consisting of four 14-cm robotic telescopes, located at the Haleakala, Cerro Tololo, South African Astronomical Observatory (SAAO) and McDonald stations of the \textcolor{black}{Las Cumbres Observatory (LCO\footnote{formerly LCOGT})}; and in \textcolor{black}{Chile}. Together, these telescopes are able to observe the entire night sky. The MASTER GLOBAL Robotic Net is a Russian collaboration of robotic telescopes distributed across the globe whose goal is the observation of the entire sky each night up to \mbox{20-21} mag with the aim of answering questions about Gamma Ray Bursts (GRBs), dark energy and exoplanets. CRTS is a transient survey, covering 33,000 square degrees of the sky between -80\textdegree and 70\textdegree \ declination, whose main goal is the discovery of rare and interesting transients. With all data being \textcolor{black}{publicly accessible}, CRTS provides valuable long-term light curves for many sources.\\
In this paper Section \ref{sec:sec2} summarises the observations and data reduction. Sections \ref{sec:sec3} - \ref{sec:sec6} contain the results of each individual CV, grouped by type: eclipsing systems (Section \ref{sec:sec3}), non-eclipsing systems in quiescence (Section \ref{sec:sec4}); non-eclipsing CVs in outburst (Section \ref{sec:sec5}), and CVs for which no period could be determined (Section \ref{sec:sec6}). Section \ref{sec:sec7} contains a summary of the data and discussion of the results.

\section{Observations} \label{sec:sec2}
Most photometric observations presented in this paper were obtained at the SAAO site in Sutherland. Differential photometry was performed using the Sutherland High-speed Optical Camera (SHOC; see \citealt{2011PASP..123..461G}; \citealt{2013PASP..125..976C}) mounted on the 74- and 40-in reflector telescopes of the SAAO. Additional observations were obtained with SHOC mounted on the SAAO's new 1-m telescope, \textcolor{black}{Lesedi \citep{lesedi_prep}}. Making use of a frame-transfer, thermoelectrically-cooled, back-illuminated CCD, SHOC allows for high-quality, high-speed photometry. \textcolor{black}{A dead time of 6.7 ms and sub-second exposure times} makes SHOC an ideal instrument to use in the search of short periods in brighter objects. This includes searching for Dwarf Nova Oscillations (DNOs, with a range of 5 - 40s), longer period DNOs (lpDNOs, with approximately 3 - 5 times the period of DNOs) and Quasi-Periodic Oscillations (QPOs, with a range of 50 -1000s) during outburst \citep{2004ASPC..310..382W}. Observations were taken with 1 MHz readout, in conventional mode \textcolor{black}{(electron-multiplying not enabled)}, with exposure times ranging from 1 to 120 seconds.\\
No filters were used and data were calibrated using PSF\_$r$ magnitudes\footnote{consistent with SDSS $r$} from either SkyMapper \citep{2007PASA...24....1K} or PANSTARRS \citep{2002SPIE.4836..154K}), unless stated otherwise. The catalogue magnitudes used to calibrate each system are given in Table \ref{tab:obstable}. As found by \cite{2014MNRAS.437..510C}, $r$-band is a close approximation to white light (WL - no or clear filter) for bluer sources ($g - r$ = 0.2 to 1.0). Since CVs are typically blue sources, we can use the $r$ magnitude as an estimate of the WL magnitude; this calibration is good to $\sim$ 0.1 mag. The data were reduced using standard \textsc{iraf} packages (such as the \verb|phot| and \verb|mkapfile| commands) to perform aperture-corrected photometry. Frequency spectrum analysis of the data was done using \textsc{eagle}, a program written by Darragh O'Donoghue for time-series analysis of unevenly spaced data containing large data gaps. \textcolor{black}{A phase dispersion minimisation \citep{1978ApJ...224..953S}, in which the sum of the variance in each bin for the folded light curve is minimised, was used to verify the periods found in eclipsing system and calculate the eclipse times.} For MLS 0720+17, photometric observations were taken with an Andor camera on the 1.3m telescope at MDM Observatory on Kitt Peak, Arizona. The MDM observations were taken with a GG420 filter, which suppresses light with wavelengths $\textless$ 4200 \AA. A log of all observations is presented in Table \ref{tab:obstable}. Only the first 10 lines are shown here, with the full version available \href{}{online}.\\
For two sources, ASASSN-15kw and MLS 0720+17, we include
time-series spectroscopy from the 2.4m Hiltner telescope at MDM 
Observatory on Kitt Peak, Arizona, USA.   We used the 
`modspec' spectrograph.\footnote{\color{blue}\url{http://mdm.kpno.noao.edu/Manuals/ModSpec/modspec_man.html}}. 
For ASASSN-15kw, the CCD covered from 4340 to 7500 \AA\ at
\textcolor{black}{3.5 \AA\ resolution} FWHM, with severe vignetting toward
the red end of the range. For MLS 0720+17, a $2048 \times 2048$ SITe CCD that covered 4210 to 7500 \AA\ with 3.6 \AA\ resolution was used. MLS 0720+17 set early in the night, limiting the observable hour angle range.
The calibration, reduction, and analysis protocols followed the same steps described by \cite{2016AJ....152..226T} and \cite{2013AJ....146..107T}. Table \ref{tab:specjournal} gives
a journal of these observations.

\begin{table*}
\caption{Observing log. Only the first 10 lines are shown here. The full version is available \href{}{online}.}
\label{tab:obstable}
\begin{tabular}{lclcccccc}
\hline
\\
Object & Type & Run & Telescope & Date of obs. & HJD of start of run & Length & $t_{in}$  & $r$\\
& & & &  (start of night) & (+2450000) & (hrs) & (s) & (mag)\\
\\
\hline
\\
\textbf{ASASSN-14eq} & SU & S8508 & 74-in & 2014/11/19 & 6981.3886 & 2.58 & 60 & 15.6- 18.5$^1$\\
& & S8515 & 74-in & 2014/11/23 & 6985.3277 & 2.83 & 30 & \\
& & S8516 & 74-in & 2014/11/24 & 6986.3093 & 1.00 & 10 & \\
& & S8517 & 74-in & 2014/11/25 & 6987.2580 & 3.72 & 10 & \\
& & S8733 & 40-in & 2015/08/05 & 7240.4354 & 6.21 & 10 & \\
& & S8735 & 40-in & 2015/08/06 & 7241.4772 & 5.13 & 10 & \\
& & S8737 & 40-in & 2015/08/07 & 7242.4807 & 5.14 & 10 & \\
& & S8739 & 40-in & 2015/08/08 & 7243.4939 & 4.79 & 10 & \\
& & S8741 & 40-in & 2015/08/09 & 7244.4304 & 1.05 & 10 & \\
\textbf{ASASSN-14hq} & DN & S8496 & 74-in & 2014/11/13 & 6975.5570 & 0.97 & 20 & 18.8 - 21.7$^1$\\
\\
\hline
\multicolumn{9}{p{15cm}}{Notes: $t_{in}$: integration time; DN: dwarf nova; SU: SU Ursae Majoris; IP: intermediate polar;
$^*$: system was in outburst; $r$: $r$ magnitude of the system in quiescence;
$^1$ makes use of the SkyMapper catalogue \citep{2018PASA...35...10W} for magnitude calibration.}
\end{tabular}
\end{table*}

\begin{table*}
\centering
\caption{Spectroscopic Journal for ASASSN-15kw and MLS 0720+17.}
\label{tab:specjournal}
\begin{tabular}{lccccc}
\hline
\\
Object & \textcolor{black}{Type} & \textcolor{black}{Date of obs}. & \textcolor{black}{HJD of start of run} & \textcolor{black}{Length} & \textcolor{black}{$t_{in}$}\\
& &  (start of night) & (+2450000) & (hrs) & (s)\\
\\
\hline
\\
\textbf{ASASSN-15kw} & DN & 2017/04/01 & 7845.0411 &  0.5  & 600\\
& & 2017/04/02 & 7845.8097 & 3.0  & 600\\
& & 2017/04/03 & 7846.8028 &  0.5  & 600\\
\textbf{MLS 0720+17} & SW Sex & 2015/04/28 & 7140.6566 &  1.0  & 900\\
& & 2015/04/29 & 7141.6448 &  1.5  & 900\\
& & 2015/04/30 & 7142.6388 &  1.5  & 900\\
\\
\hline
\end{tabular}
\end{table*}

\section{Eclipsing systems} \label{sec:sec3}
\begin{figure*}
\centering
\makebox[\textwidth][c]{\includegraphics[width=\columnwidth]{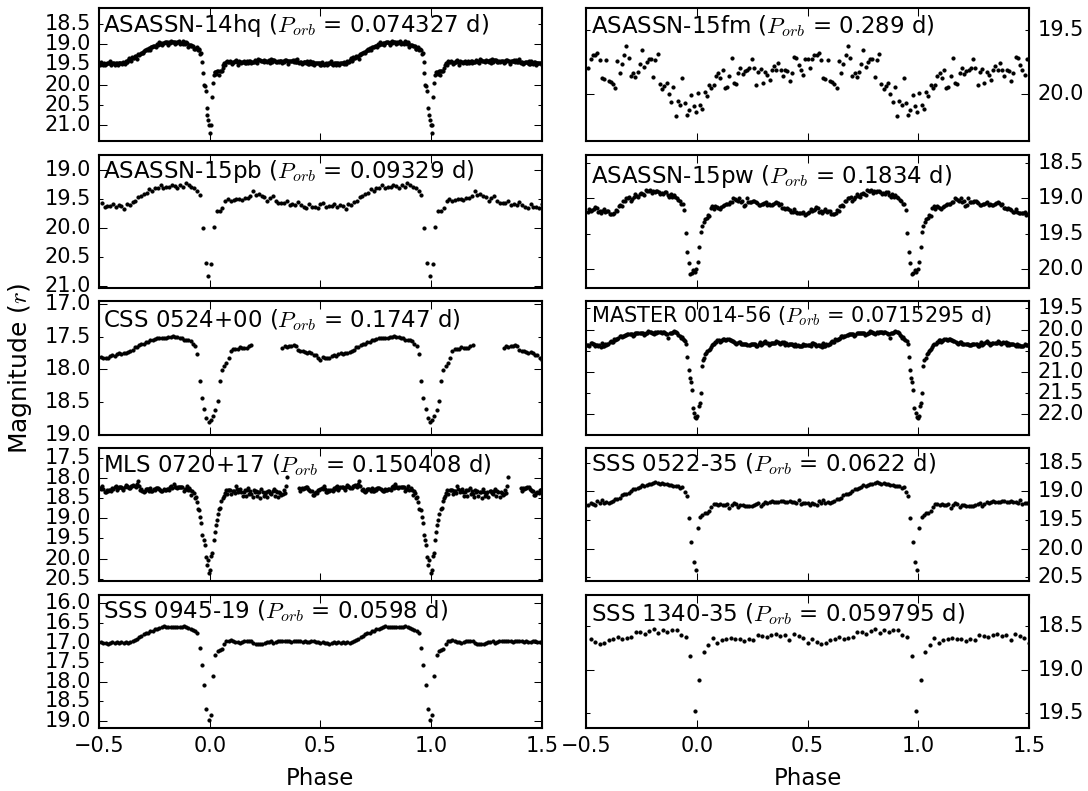}}
\caption{Our average light curves of eclipsing systems presented in this paper, duplicated over 2 orbital cycles. The system's name and orbital period is shown in the plot. For ASASSN-14hq, runs S8825 and S8831 are excluded due to bad weather, and the target being in a brighter state (possibly on the rise of a normal outburst) respectively. Due to lack of data, phase 0.19-0.3 is not plotted for CSS 0524+00. Due to lack of data, phase 0.35-0.4 is not plotted for MLS 0720+17 and only runs obtained in Sutherland, \textcolor{black}{denoted with run names containing 'S'}, are included. \textcolor{black}{Many of these systems show strong orbital humps in their light curves.}}
\label{fig:eclipsing}
\end{figure*}

This section contains the details of the eclipsing systems presented in this paper. These systems are listed in alphabetical order and our average light curves, except for ASASSN-14ka, are shown in Figure \ref{fig:eclipsing}. Eclipsing systems play an important role in the study of CV evolution through the modeling of eclipse profiles \citep{2017MNRAS.465.4968H}. With the exception of ASASSN-14ka and ASASSN-15fm which have shallow eclipses, the eclipsing systems in this paper show narrow eclipses, ranging from 0.2 to 2 mag in depth. \textcolor{black}{A table listing eclipse times for each system presented in this paper is available \href{}{online}.}

\subsection{ASASSN-14hq}
ASASSN-14hq shows evidence of previous outbursts, as well as eclipses, in the CRTS data. It was identified as a CV candidate by the ASAS-SN survey on 2014 September 24, when it went into outburst reaching $V = 13.97$ mag \textcolor{black}{\citep{2014AAS...22323603S}}. ASASSN-14hq shows a characteristic light curve of an eclipsing dwarf nova in quiescence. Our average light curve is shown in Figure \ref{fig:eclipsing}, and shows deep, narrow eclipses of more than 1 mag in depth during quiescence. ASASSN-14hq has an orbital period of $0.074327(\pm9)$ d and the eclipse ephemeris is
\begin {equation}\label{eq:14hq}
\mathrm{HJD_{min}} = 2456975.5967(\pm2) + 0.074327(\pm9)\ E
\end{equation}
With an orbital period below the period gap, we expect this system to have superoutbursts. Although the archival data of CRTS does show outbursts, there is inadequate coverage to determine whether these are normal outbursts, or in fact superoutbursts. Archival data from ASAS-SN however, show evidence of regularly occurring normal outbursts and superoutbursts.

\subsection{ASASSN-14ka}
\begin{figure*}
\centering
\includegraphics[width=\columnwidth]{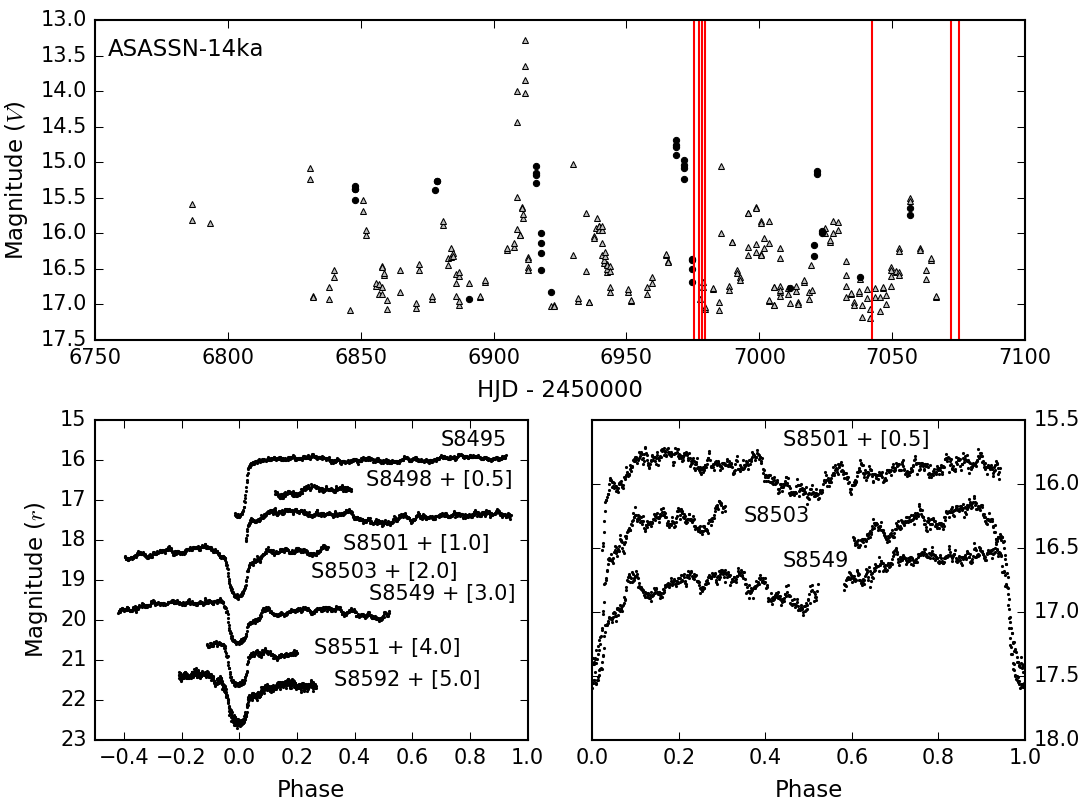}
\caption{Top: Long term light curve of ASASSN-14ka obtained from the ASAS-SN survey team. The grey triangle indicate upper limits, while the red lines show when the observation presented in this paper where taken. Bottom left: Individual light curves of ASASSN-14ka. The light curve for run S8495 is displayed at the correct brightness; the vertical offset for each light curve thereafter, is given in brackets. The figure clearly shows the structure present in quiescence, as well as the changing eclipse profile as the system was declining from outburst. Bottom right: Individual light curves from runs S8501, S8503 and S8549, folded on the ephemeris given in equation \ref{eq:14ka}. Run S8501 vertically offset by 0.5 mag for display purposes. This plot highlights the modulation seen in ASASSN-14ka at half the orbital period.}
\label{fig:14ka}
\end{figure*}

ASASSN-14ka, was announced as a CV candidate by the ASAS-SN survey on 2014 September 15, when it underwent an outburst peaking at $V = 15.06$ mag \textcolor{black}{\citep{2014AAS...22323603S}}. It was also reported by the Gaia Photometric Science Alerts \citep{2012gfss.conf...21W} in 2017 as Gaia17anx. Archival data from the ASAS-SN survey team \citep{2014AAS...22323603S}, displayed in Figure \ref{fig:14ka}, show regular outbursts occurring approximately once a month. Our SHOC light curves obtained are displayed in Figure \ref{fig:14ka}, each being vertically offset for display purposes. During runs S8495, S8498, S8501 and S8503, the system was still in a brightened state, with a resulting eclipse minimum of 18.4 mag. During the three later runs (S8549, S8551 and S8592), the system had returned to quiescence and showed a slightly deeper eclipse minimum of 18.7 mag. ASASSN-14ka has an orbital period of $0.17716(\pm1)$ d and the eclipse ephemeris is
\begin {equation}\label{eq:14ka}
\mathrm{HJD_{min}} = 2456975.3854(\pm2) + 0.17716(\pm1)\ E
\end{equation}  
Evidence of a modulation with an amplitude of 0.6 magnitude at half the orbital period can be seen in the bottom right panel of Figure \ref{fig:14ka}. While ASASSN-14ka shows flickering on the order of 0.2 mag, no evidence of other periods was found.

\subsection{ASASSN-15fm}
\begin{figure}
\includegraphics[width=\columnwidth]{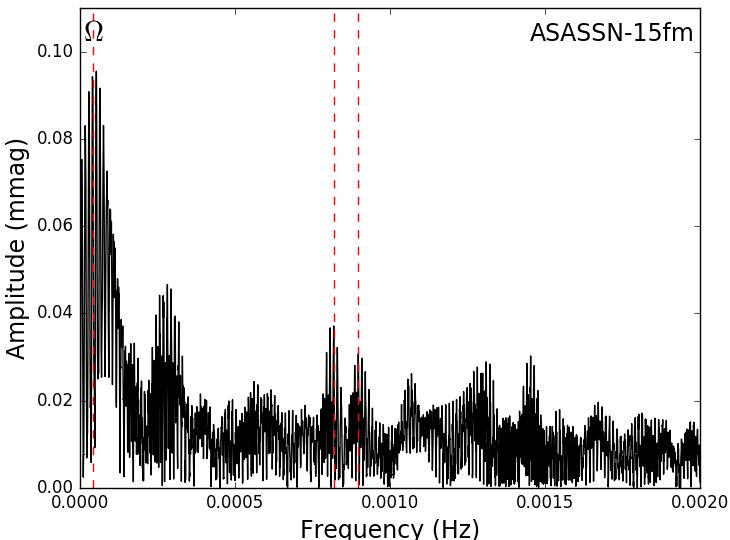}
\caption{FT of three longest runs (S8638, S8640, S8643) \textcolor{black}{for ASASSN-15fm}. The dashed lines show the orbital period ($\Omega$), the spin of the white dwarf, and a 2nd side band.}
\label{fig:15fm1}
\end{figure}

ASASSN-15fm, was announced as a CV candidate by the ASAS-SN survey on 2015 March 15, when it went into outburst with a peak magnitude of $V = 16.26$ mag \textcolor{black}{\citep{2014AAS...22323603S}}. The Fourier Transform (FT) of the three longest runs, displayed in Figure \ref{fig:15fm1}, shows that ASASSN-15fm is a probable intermediate polar (IP). The highest peak in the FT is most likely the orbital period (represented by $\Omega$), while the spin of the white dwarf, along with the interaction between these two periods, appears as the two smaller peaks highlighted by the dashed lines. The two smaller periods at 18.57$(\pm1)$ and 20.37$(\pm1)$ mins are separated by twice the orbital frequency, but the duration of our data is insufficient to distinguish which is the spin period of the white dwarf. A more detailed study at higher cadence is needed to determine the spin period of the white dwarf. Our average light curve is shown in Figure \ref{fig:eclipsing}. The orbital ephemeris for minimum light is
\begin {equation}\label{eq:15fm}
\mathrm{HJD_{min}} = 2457134.708(\pm5) + 0.289(\pm1)\ E
\end{equation}

\subsection{ASASSN-15pb}
The CRTS light curve provides no evidence of previous outbursts or eclipses for ASASSN-15pb due to sparse coverage. ASASSN-15pb was listed as a CV candidate by the ASAS-SN survey on 2015 September 1, when it went into outburst with $V = 16.48$ mag \textcolor{black}{\citep{2014AAS...22323603S}}. Our average light curve is shown in Figure \ref{fig:eclipsing}, and shows eclipse depths of more than 1.5 mag in quiescence. ASASSN-15pb has an orbital period of $0.09329(\pm2)$ d, just above the period gap minimum of $2.15\textcolor{black}{(\pm0.03)}$ hrs \citep{2011ApJS..194...28K}, and the eclipse ephemeris is
\begin {equation}\label{eq:15pb}
\mathrm{HJD_{min}} = 2457312.349(\pm1) + 0.09329(\pm2)\ E
\end{equation}

\subsection{ASASSN-15pw}
ASASSN-15pw shows evidence of previous outbursts, but no eclipses, within the CRTS data. It was listed as a CV candidate by the ASAS-SN survey on 2015 September 22, when it went into outburst with $V = 16.07$ mag \textcolor{black}{\citep{2014AAS...22323603S}}. Our average light curve is shown in Figure \ref{fig:eclipsing}, and shows eclipses with a depth of around 1 mag. ASASSN-15pw has an orbital period of $0.1834(\pm3)$ d and the eclipse ephemeris is
\begin{equation}\label{eq:15pw}
\mathrm{HJD_{min}} = 2457316.589(\pm1) + 0.1834(\pm3)\ E
\end{equation}

\subsection{CSS 0524+00 (CSS131106:052412+004148)}
Since the discovery of CSS 0524+00 by CRTS on 2013 Nov 6 \textcolor{black}{\citep{2009ApJ...696..870D}}, ample coverage shows evidence of multiple eclipses and outbursts. With evidence of eclipses, \cite{2017MNRAS.465.4968H} observed CSS 0524+00, finding a period of 0.17466647$(\pm2)$ d. We find a period of 0.1747$(\pm3)$ d, in agreement with that found by \cite{2017MNRAS.465.4968H}. Our average light curve, folded on the ephemeris $\mathrm{HJD_{min}}  = 2456651.4295(\pm3) + 0.1747(\pm3)\ E$, is shown in Figure \ref{fig:eclipsing}; and shows eclipse depths of around 1.2 mag.

\subsection{MASTER 0014--56 (MASTER OT J001400.25--561735.0)}
CRTS data of MASTER 0014--56 show evidence of eclipses, along with a possible outburst. MASTER 0014--56 was discovered by MASTER-SAAO when it went into outburst with an amplitude of more than 3.7 mag \citep{2015ATel.7860....1G}. Our average light curve is shown in Figure \ref{fig:eclipsing}, and shows deep, narrow eclipses of $\sim$ 2 mag depth in quiescence. MASTER 0014--56 has an orbital period of $0.0715295(\pm6)$ d and the eclipse ephemeris is
\begin {equation}\label{eq:00-56}
\mathrm{HJD_{min}} = 2457245.5459(\pm1) + 0.0715295(\pm6)\ E
\end{equation}

\subsection{MLS 0720+17 (MLS101226:072033+172437)}
\begin{figure}
\includegraphics[width=\columnwidth]{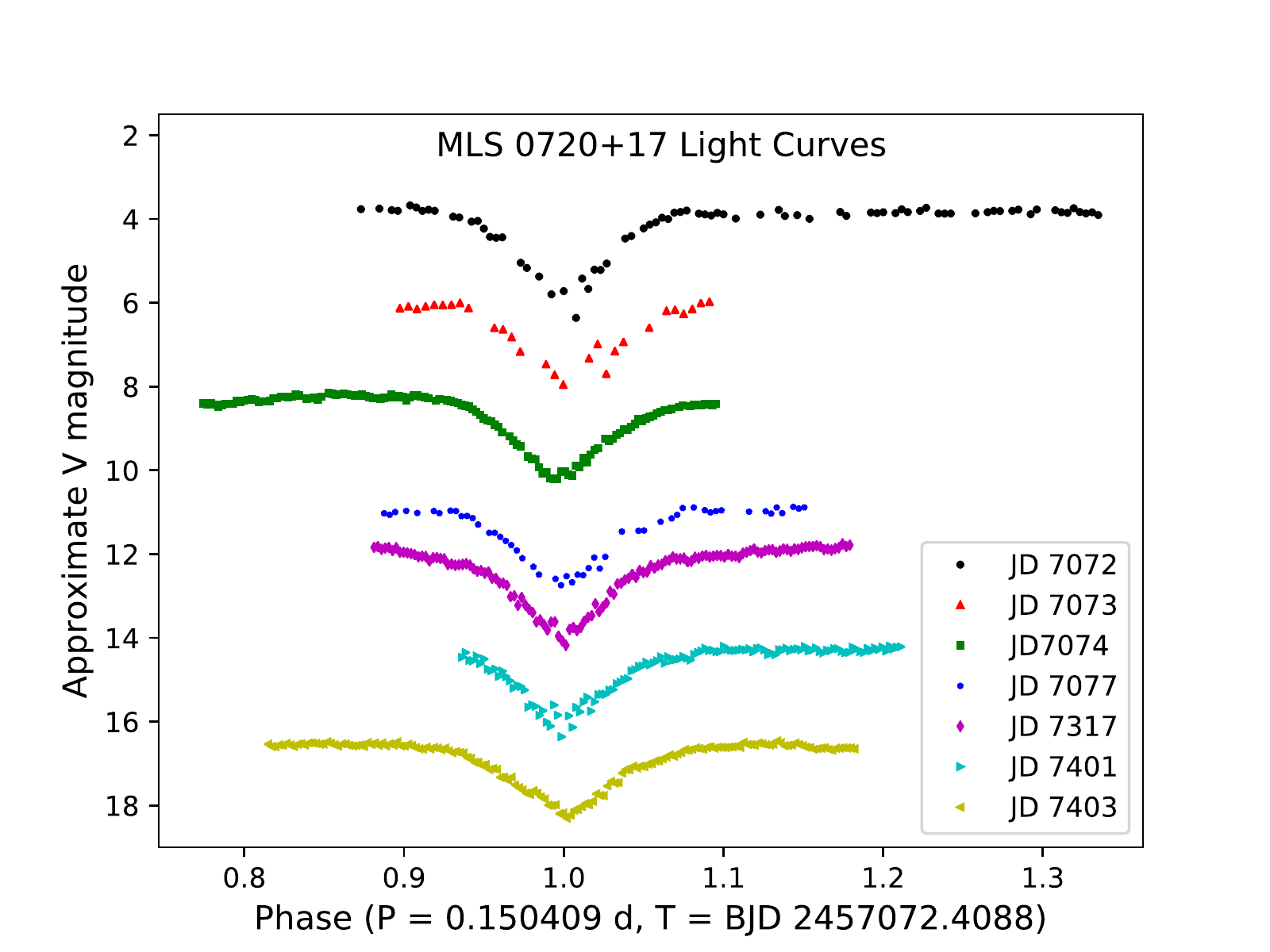}
\caption{Differential photometry in $V$ \textcolor{black}{(approximated by the shift in white light by the comparison star)} of MLS 0720+17 as a function
of orbital phase. Each light curve is offset by $\Delta V = 3$ respectively.
The first four nights' data were taken with SHOC without a
filter. The V-shaped eclipse is another trait of SW
Sex stars.}
\label{fig:20+17_photo}
\end{figure}

\begin{figure}
\includegraphics[width=\columnwidth]{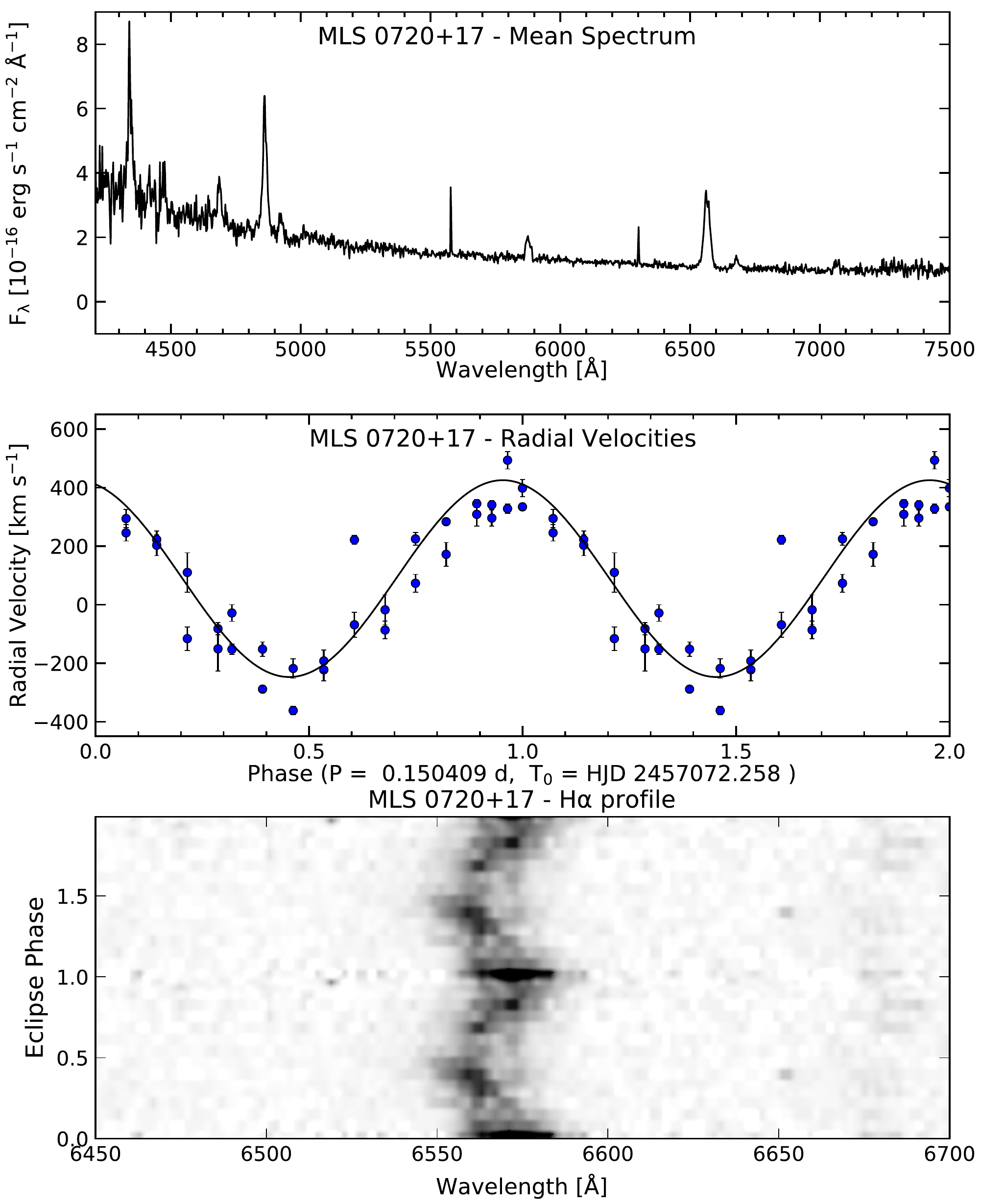}
\caption{(Upper) Mean spectrum of MLS 0720+17 from MDM
data taken April 2017. (Mid) $H_{\alpha}$ radial velocities folded
on the eclipse ephemeris given in equation \ref{eq:20+17}.
There is an apparent 0.175($\pm$0.031) cycle phase shift.
(Lower) $H_{\alpha}$ plotted as a function of phase.}
\label{fig:20+17_spec}
\end{figure}

After the discovery of MLS 0720+17 by CRTS, \cite{2009ApJ...696..870D} interpreted the variability seen in the light curve as eclipses. \cite{2017MNRAS.465.4968H} confirmed the presence of eclipses when they obtained a short observation, in which they observed part of an eclipse. \cite{2017AJ....153..144O} later obtained a spectrum of MLS 0720+17. They concluded that the spectrum was typical of a polar, with the eclipse seen by \cite{2017MNRAS.465.4968H} being a modulation due to cyclotron emission, and the narrow emission lines seen in the spectrum inconsistent with an eclipsing disc system. Our individual light curves and time-resolved spectroscopy are shown in Figure \ref{fig:20+17_photo} and Figure \ref{fig:20+17_spec} respectively, while our average light curve is shown in Figure \ref{fig:eclipsing}. Our observations confirm MLS0720+17 as an eclipsing system.
From the four eclipses in the 2015 SAAO light curves, we found a preliminary period of 0.1504 d,
which we constrained further using the 2013 eclipse from SAAO and the October and January eclipses from MDM. \textcolor{black}{No sign of a coherent pulsation, as would be expected for an IP, was seen.} The eclipse ephemeris is
\begin{equation}\label{eq:20+17}
\mathrm{HJD_{min}} = 2457072.2590(\pm7) + 0.150408(\pm7)\ E
\end{equation}
The radial velocities, \textcolor{black}{determined using the convolution method described by \cite{1980ApJ...238..946S}}, do not independently determine the period due to the limited time span, but they do show a strong modulation consistent with the known period. We fit the velocities with a sinusoid of the form
\begin{equation}\label{eq:sin}
v(t) = \gamma + K \sin (2\pi (t - t_0) / P)
\end{equation}
using linear least squares, with the period P fixed at the value derived from the eclipses. This yielded
$t_0$ = $\mathrm{HJD_{0}}\ = 2457141.698(\pm0.003)$, $K = 299(\pm23)$ km s$^{-1}$, and
$\gamma = 88(\pm$19) km s$^{-1}$.
The radial velocities, folded on the eclipse ephemeris, are shown in Figure \ref{fig:20+17_spec} with the sinusoidal fit superposed. There is a phase difference between the radial velocity fit and eclipse phase of \textcolor{black}{0.175$(\pm0.031)$} cycles, a hallmark of a subclass of CVs, the SW Sextantis (SW Sex) stars. First classified by \cite{1991AJ....102..272T}. These are nova-like CV stars that exhibit a suite of properties as follows: (1) Absorption in the Balmer and \ion{He}{I} lines appears near orbital phase 0.5; (2) An S-wave absorption feature in the emission of $H_{\alpha}$ is often observed; (3) In cases in which the true orbital phase is known from eclipses, the zero phase of the radial
velocities lag behind the eclipses if they were to trace the white dwarf's motion; in other words, at eclipse, the Balmer line velocities have not yet decreased to their mean value; (4) The orbital periods of SW Sex stars are clustered from the 3 hour upper limit of the period gap up to about 4 hours \citep{2009MNRAS.395..973R}.

In many SW Sex stars, the HeI and Balmer absorption features appear around phase 0.5, opposite the eclipse. This is not apparent in the present data, most likely due to low signal-to-noise. Further studies are necessary to determine whether the absorption feature appears. \textcolor{black}{The spectrum seems to show 2 additional traits of SW Sex stars: signs of a HeII 4686 + Bowen blend, and singly peaked emission lines.} The bottom panel of Figure \ref{fig:20+17_spec} shows a grey-scale representation of the $H_{\alpha}$ line as a function of phase. The velocity shifts are readily apparent. There is an artificial brightening of the intensity of the $H_{\alpha}$ line during eclipse (phase 0 and 1) because the line is normalized to the continuum, boosting the line when the continuum is eclipsed. SW Sex stars' eclipses are often deeper in the continuum than in the lines \citep{1998ASPC..137..132D}. \cite{2001A&A...368..183G} showed that this effect is a result of the emission lines forming above the disc. 

\subsection{SSS 0522--35 (SSS111126:052210--350530)}
SSS 0522--35 was discovered by CRTS on 2011 Nov 11, with a peak outburst amplitude of 2.53 mag \textcolor{black}{\citep{2009ApJ...696..870D}}. CRTS data show evidence of high variability and outbursts roughly 5-6 months apart. Our average light curve is shown in Figure \ref{fig:eclipsing}, and shows eclipse depths of around 1.5 mag. SSS 0522--35 has an orbital period of $0.0622(\pm5)$ d and the eclipse ephemeris is
\begin{equation} \label{eq:05-35}
\mathrm{HJD_{min}} = 2455913.4377(\pm2) +   0.0622(\pm5)\ E
\end{equation}

\subsection{SSS 0945--19 (SSS130413:094551--194402)}
Suspected as a variable by \cite{1981CSVS..C......0K} (known as NSV4618), the CRTS light curve of SSS 0945--19 shows evidence of deep eclipses, as well as previous outbursts. Eclipses and an orbital period of 0.065769264$(\pm2)$ d was reported by Kato, T. through the \textit{vsnet} collaboration (vsnet-alert 15615). \cite{2017MNRAS.465.4968H} observed a single eclipse, showing it to have clear white dwarf and bright spot features. Our average light curve is shown in \text{Fig. \ref{fig:eclipsing}}. SSS 0945--19 has an orbital period of $0.0657693(\pm3)$ d, in agreement with the period reported by Kato, T., and the eclipse ephemeris is
\begin{equation} \label{eq:09-19}
\mathrm{HJD_{min}} = 2456421.3609(\pm1) +  0.0657693(\pm3)\ E
\end{equation}

\subsection{SSS 1340--35 (SSS120402:134015--350512)}
After its discovery by CRTS \textcolor{black}{\citep{2009ApJ...696..870D}}, and first observed by \cite{2014MNRAS.437..510C}, SSS 1340--35 was found to be eclipsing with an orbital period of $0.059(\pm1$) d. With our new observations, the orbital period has been refined to be $0.0598(\pm1)$ d. Our average light curve is shown in \text{Fig. \ref{fig:eclipsing}}. The eclipse ephemeris is
\begin{equation} \label{eq:13-35}
\mathrm{HJD_{min}} = 2457073.5424(\pm8) + 0.0598(\pm1)\ E
\end{equation}

\section{Non-eclipsing systems in quiescence} \label{sec:sec4}
\begin{figure*}
\includegraphics[width=\columnwidth]{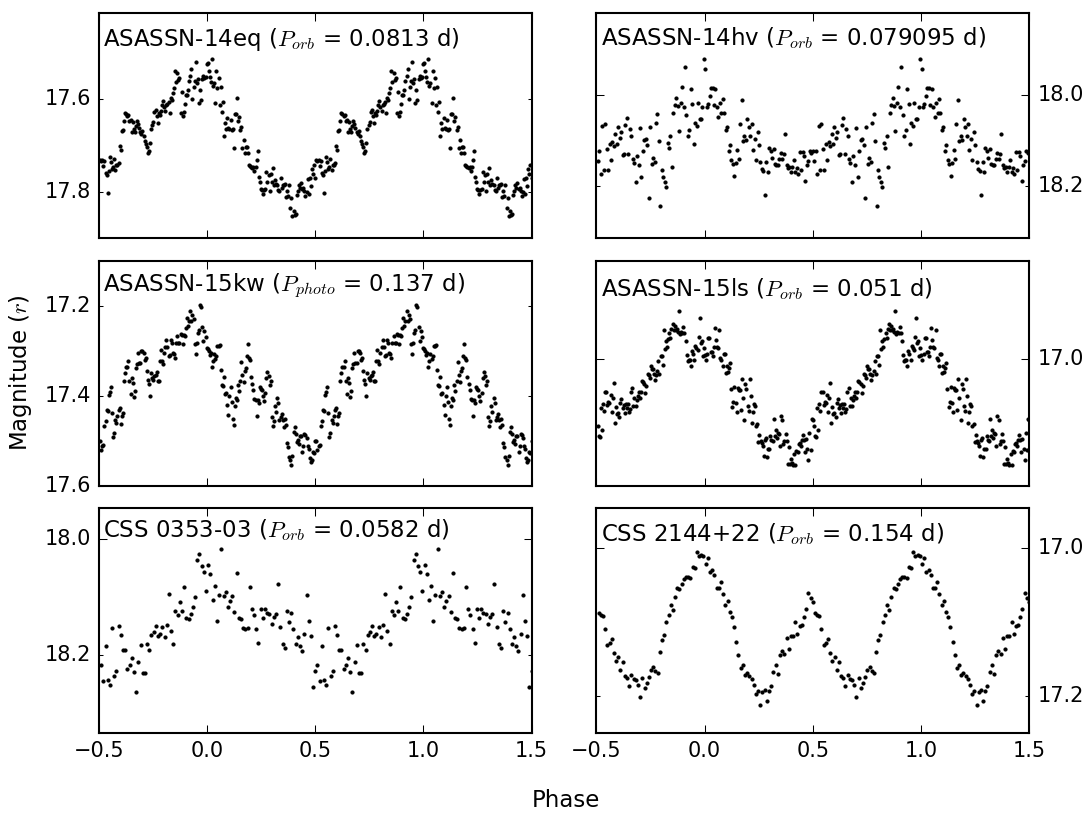}
\caption{Our average light curves of non-eclipsing systems in quiescence presented in this paper, duplicated over 2 orbital cycles. The system name and orbital period are shown in the plot. Only the later five runs (S8533 - S8545) are shown for ASASSN-14eq.}
\label{fig:non-eclipsing}
\end{figure*}

This section contains the details of individual non-eclipsing systems, for which orbital periods were found. These systems are listed in alphabetical order and our average light curves are shown in Figure \ref{fig:non-eclipsing}.

\subsection{ASASSN-14eq}
ASASSN-14eq appears in both the CRTS, as well as the All Sky Automated Survey \textcolor{black}{release 3} (ASAS-3; \textcolor{black}{\citealt{2004AcA....54..153P}}). The combined survey light curves show evidence of previous outbursts, some of which resemble superoutbursts. It was announced as a CV candidate by the ASAS-SN survey on 2014 July 28, when it underwent an outburst reaching a of $V = 13.53$ mag \textcolor{black}{\citep{2014AAS...22323603S}}. Our average light curve is shown in Figure \ref{fig:non-eclipsing}. The observations presented in this paper were taken: 1) nearly 4 months (S8508 - S8517); and 2) over a year (S8533 - S8545), after the outburst recorded by \cite{2015PASJ...67..105K}. The system was in quiescence during all of these observations. \cite{2015PASJ...67..105K} found a superhump period of 0.079467 d. Observations of ASASSN-14eq show an orbital period of $0.0813(\pm3)$ d. This results in a negative superhump excess of --0.02, consistent with the expected value for such an orbital period (\citealt{2001cvs..book.....H}, Fig. 6.19). The orbital ephemeris for maximum light is 
\begin {equation}\label{eq:14eq}
\mathrm{HJD_{max}} = 2457240.5088(\pm1) + 0.0813(\pm3)\ E
\end{equation}
ASASSN-14eq shows flickering on the order of 0.1 mag, with no evidence of other periods found.

\subsection{ASASSN-14hv}
\begin{figure}
\centering
\includegraphics[width=\columnwidth]{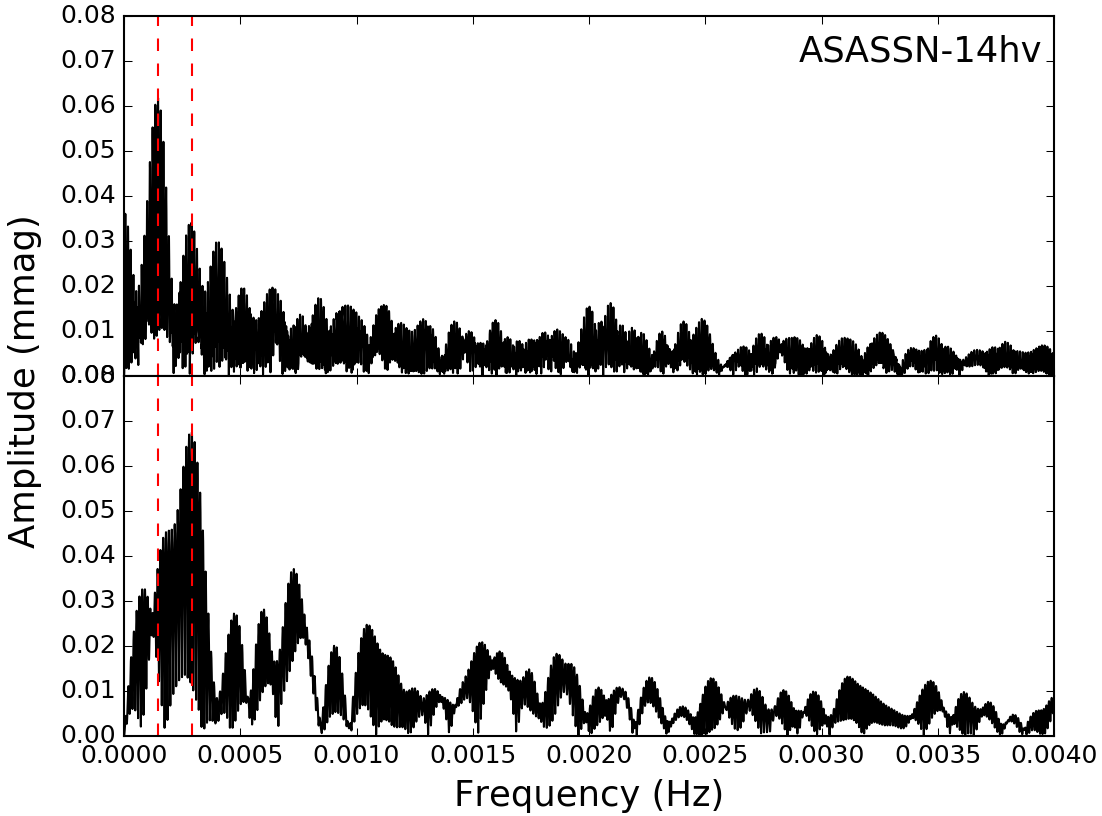}
\caption{Above: FT of runs S8509, S8511, S8513 and S8514. Below: FT of runs S8851 and S8854. The dashed lines show the orbital period and the \textcolor{black}{first} harmonic. It is interesting to note the shifting of power between the orbital period and the harmonic. The reason for this is unclear.}
\label{fig:14hv2}
\end{figure}

ASASSN-14hv was announced as a CV candidate by the ASAS-SN survey on 2014 September 27, when it underwent an outburst with a peak of $V = 14.16$ mag \textcolor{black}{\citep{2014AAS...22323603S}}, and shows outbursts, including superoutbursts, approximately once every 2 months. Our observations, excluding runs S8846 and S8847 in which the system appear to be declining from outburst, were taken when the system was in quiescence. From run S8847, we determine a superhump period of $0.082(\pm2)$ days. Our average light curve in quiescence is shown in Figure \ref{fig:non-eclipsing}. With an orbital period of $0.079095(\pm7)$ d, ASASSN-14hv shows a superhump excess of 0.04 consistent with the expected value \citep{2001cvs..book.....H}. The FT of the earlier runs (from Nov 2014) and the later runs (from May 2017) are shown in Figure \ref{fig:14hv2}. The orbital period and its harmonic can be seen in the FTs, but with the power changing between the orbital period and the harmonic. The orbital ephemeris for maximum light is
\begin {equation}\label{eq:14hv}
\mathrm{HJD_{max}} = 2456981.6068(\pm7) + 0.079095(\pm7)\ E
\end{equation}

\subsection{ASASSN-15kw}
\begin{figure}
\centering
\includegraphics[width=\columnwidth]{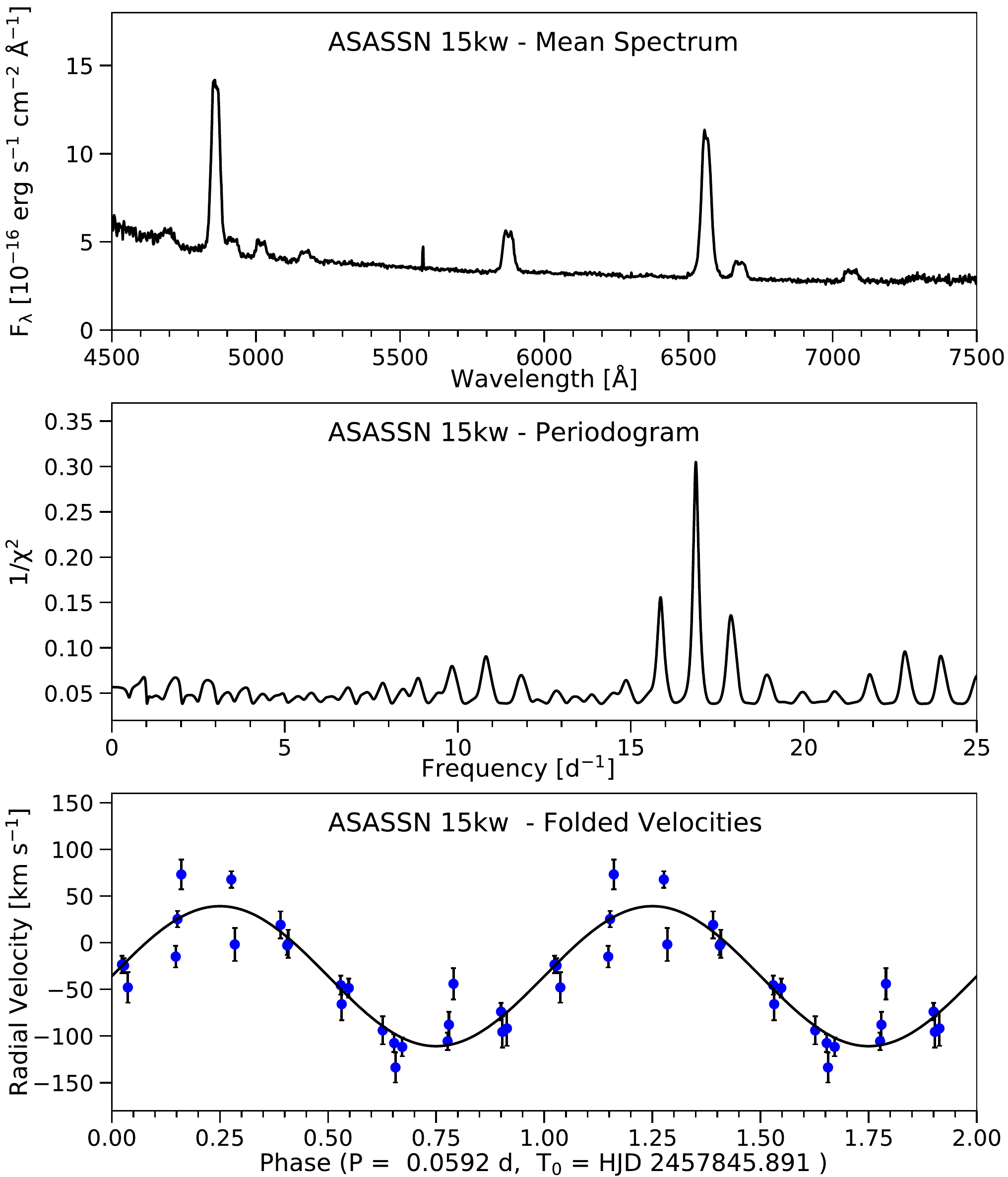}
\caption{(Upper) Mean spectrum of ASASSN-15kw from MDM
data taken April 2017. (Mid) Periodogram of the H$\alpha$
emission velocities.  The peaks flanking the main peak
are aliases caused by the sampling.  (Lower) Radial
velocities of H$\alpha$, plotted across 2 phases, 
folded on the best-fitting spectroscopic ephemeris,
with the best-fit sinusoid superposed.  Uncertainties
shown are derived from counting statistics.}
\label{fig:specmontage}
\end{figure}

With a possible outburst in the CRTS, ASASSN-15kw was announced as a CV candidate by the ASAS-SN survey on 2015 June 10, when it went into outburst with a peak magnitude of $V = 14.44$ mag \textcolor{black}{\citep{2014AAS...22323603S}}. \cite{2015ATel.7641....1C} obtained a spectrum of ASASSN-15kw, confirming its classification as a CV. Our average light curve is shown in Figure \ref{fig:non-eclipsing}. \textcolor{black}{Individual light curves} show flickering with an amplitude on the order of 0.3 mag. The ephemeris for maximum light is
\begin {equation}\label{eq:15kw}
\mathrm{HJD_{max}} = 2457217.3939(\pm1) + 0.137(\pm6)\ E
\end{equation}
Our mean spectrum, shown in Figure \ref{fig:specmontage}, is typical
of dwarf novae at minimum light.  
The emission lines are almost double-peaked, with typical 
FWHM near 2000 km s$^{-1}$; the emission equivalent widths
of H$\beta$ and H$\alpha$ are respectively --80 and --115 \AA .
The H$\alpha$ radial velocities show a significant modulation
at an unambiguously-determined period of 0.05924$(\pm10)$ d, or 85.3 min; 
a sinusoidal fit in the form of equation \ref{eq:sin}
gives $t_0 = 2 457 845.891(\pm1) {\rm\ BJD}$, 
$\gamma = -36(\pm6)$ km s$^{-1}$, and $K = 75(\pm9)$ km s$^{-1}$ at this period.

It is likely that the 85-min period is $P_{\rm orb}$, and the
3.28-hour photometric period arises from some other 
phenomenon.  We do not have a ready explanation for these
two periods, but note that \cite{2002Ap&SS.282..433W} found a 
similar discrepancy in another short-period dwarf nova, 
GW Lib; its orbital period is 1.28 h, but they found a
significant photometric modulation at 2.09 h, and they note
similar discrepant periodicities in FS Aur and V2051 Oph.  

\subsection{ASASSN-15ls}
ASASSN-15ls, not covered by the CRTS, was announced as a CV candidate by the ASAS-SN survey on 2015 June 19, when it went into outburst with a peak magnitude of $V = 16.33$ mag \textcolor{black}{\citep{2014AAS...22323603S}}. Our average light curve is shown in Figure \ref{fig:non-eclipsing}. The orbital ephemeris for maximum light is
\begin {equation}\label{eq:15ls}
\mathrm{HJD_{max}} = 2457240.2242(\pm1) +0.051(\pm8)\ E
\end{equation}

\subsection{CSS 0353-03 (CSS111231:035318-034847)}
After its discovery on 2011 Dec 31 \textcolor{black}{\citep{2009ApJ...696..870D}}, \cite{2014AJ....148...63S} obtained a spectra of CSS 0353-03 during the 2013 Jan outburst. The spectrum showed a flat blue continuum. CRTS data show evidence of outbursts occurring roughly once a year. Our average light curve is shown in Figure \ref{fig:non-eclipsing}. The orbital ephemeris for maximum light is
\begin{equation} \label{eq:03-03}
\mathrm{HJD_{max}} = 2456247.5218(\pm2) + 0.0582(\pm1)\ E
\end{equation}

\subsection{CSS 2144+22 (CSS100520:214426+222024)}
CSS 2144+22 was discovered by CRTS on 2010 May 20, with a peak outburst amplitude of 2.41 mag \textcolor{black}{\citep{2009ApJ...696..870D}}. The CRTS light curve shows evidence of previous outbursts, as well as a possible superoutburst. \cite{2017MNRAS.465.4968H} observed CSS 2144+22, confirming no eclipses. Our average light curve is shown in Figure \ref{fig:non-eclipsing}. The orbital ephemeris for maximum light is
\begin{equation} \label{eq:21+22}
\mathrm{HJD_{max}} = 2456564.3701(\pm3) + 0.154(\pm1)\ E
\end{equation}

\section{Non-eclipsing systems in outburst} \label{sec:sec5}
\begin{figure*}
\centering
\includegraphics[width=\columnwidth]{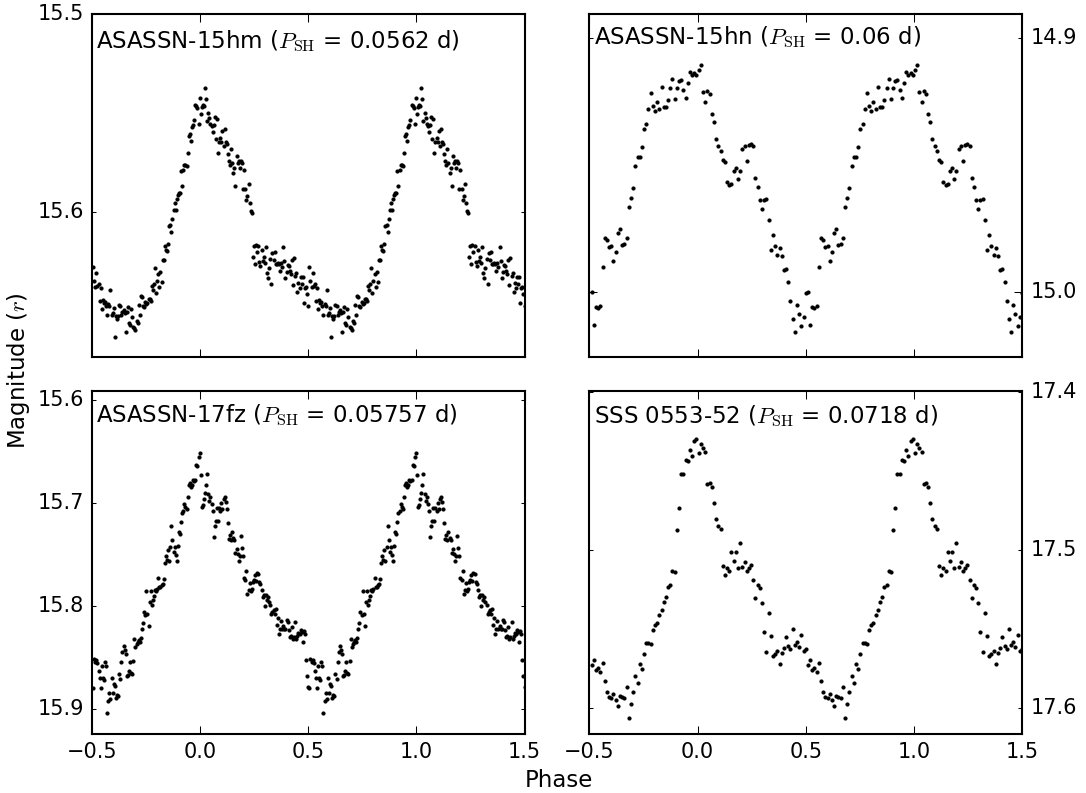}
\caption{Our average light curves of systems in outburst presented in this paper duplicated over 2 orbital cycles. The system name and superhump period is shown in the plot. Only the longest two runs (S8649 and S8651) are shown for ASASSN-15hm.}
\label{fig:outburst}
\end{figure*}

This section contains the details of individual systems which were observed during outburst, and for which superhump periods were found. Our average light curves are shown in Figure \ref{fig:outburst}. Our observations for ASASSN-17fz are shown in Figure \ref{fig:long}, to show the overall shape of the observed outburst.

\subsection{ASASSN-15hm}
ASASSN-15hm was observed during outburst after it was announced as a CV candidate by the ASAS-SN survey on 2015 April 18, \textcolor{black}{\citep{2014AAS...22323603S}}. ASASSN-15hm was also detected a month later by Gaia Photometric Science Alerts \citep{2012gfss.conf...21W} as Gaia15aeu. \cite{2015ATel.7641....1C} obtained a spectrum of ASASSN-15hm, but classified it as a stellar object due to its redness and strong, narrow \ion{Na}{D} absorption. \cite{2016PASJ...68...65K} reported a superhump period of 0.056219 d. Using the two longest runs (as these were the cleanest with multiple orbital cycles) presented in this paper (S8649 and S8651), a superhump period of $0.0562(\pm1)$ d was found. This period agrees with the superhump period found by \cite{2016PASJ...68...65K}. Our average light curve of run S8653, folded on the ephemeris $\mathrm{HJD_{max}}  = 2457145.22513(\pm1) + 0.0562(\pm1)\ E$, is shown in Figure \ref{fig:outburst}.

\subsection{ASASSN-15hn}
ASASSN-15hn was observed during outburst after it was announced as a CV candidate by the ASAS-SN survey on 2015 April 18, \textcolor{black}{\citep{2014AAS...22323603S}}. We obtained 4 runs, the later three runs (S8647, S8653 and S8657) overlap with the data presented by \cite{2016PASJ...68...65K}. These runs provided limited coverage, and a superhump period of 0.06$(\pm1)$ was found using them. \cite{2016PASJ...68...65K} reported a superhump period of 0.06189 d. Our average light curve of run S8653, folded on the ephemeris $\mathrm{HJD_{max}}  = 2457147.20394(\pm1) + 0.06(\pm1)\ E$, is shown in Figure \ref{fig:outburst}.

\subsection{ASASSN-17fz}
\begin{figure}
\centering
\includegraphics[width=\columnwidth]{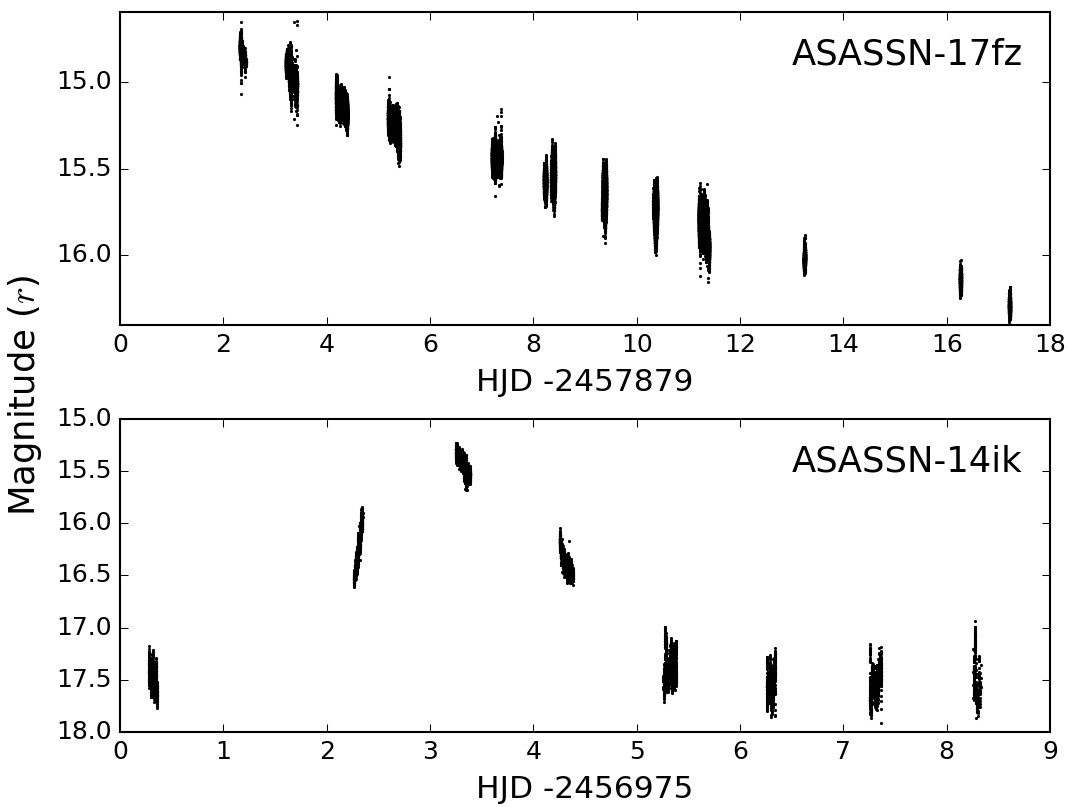}
\caption{Multiple runs showing the magnitude of systems observed in outburst. Top: All observations of ASASSN-17fz presented in this paper. Bottom: Observations of ASASSN-14ik showing the general shape of the Nov 2014 outburst.}
\label{fig:long}
\end{figure}

ASASSN-17fz was observed during a superoutburst after it was announced as a CV candidate by the ASAS-SN survey on 2017 May 5, \textcolor{black}{\citep{2014AAS...22323603S}}. Our observations of ASASSN-17fz are shown in Figure \ref{fig:long}. A single observation obtained by D. L. Holdsworth on the SAAO 1-m with SHOC on 12 Dec 2017 placed an upper limit of 21 mag on the brightness of the system. \textcolor{black}{With a quiescent magnitude of $>$21}, the amplitude of the observed superoutburst is more than six magnitudes. The superoutburst also showed a slow decline of $\sim$ 0.13 mag per day for the duration of the observing period, classifying ASASSN-17fz as a WZ Sge-type star. Our average light curve is shown in Figure \ref{fig:outburst}. A superhump period of $0.05757(\pm5)$ d was found during the outburst with an ephemeris for maximum light during outburst given by
\begin {equation}\label{eq:17fz}
\mathrm{HJD_{max}} = 2457888.40859(\pm1) + 0.05757(\pm5)\ E
\end{equation}

\subsection{SSS 0553-52 (SSS111213:055349-525045)}
Since its discovery by CRTS \textcolor{black}{\citep{2009ApJ...696..870D}} on 2011 Dec 13, there has been sparse coverage of SSS 0553-52 within CRTS. Our average light curve is shown in Figure \ref{fig:outburst}.
The orbital ephemeris for maximum light is
\begin{equation} \label{eq:05-52}
\mathrm{HJD_{max}} = 2455910.4994(\pm1) +  0.0718(\pm2)\ E
\end{equation}

\section{CVs for which no periodicity could be determined} \label{sec:sec6}
\begin{figure*}
\centering
\includegraphics[width=\columnwidth]{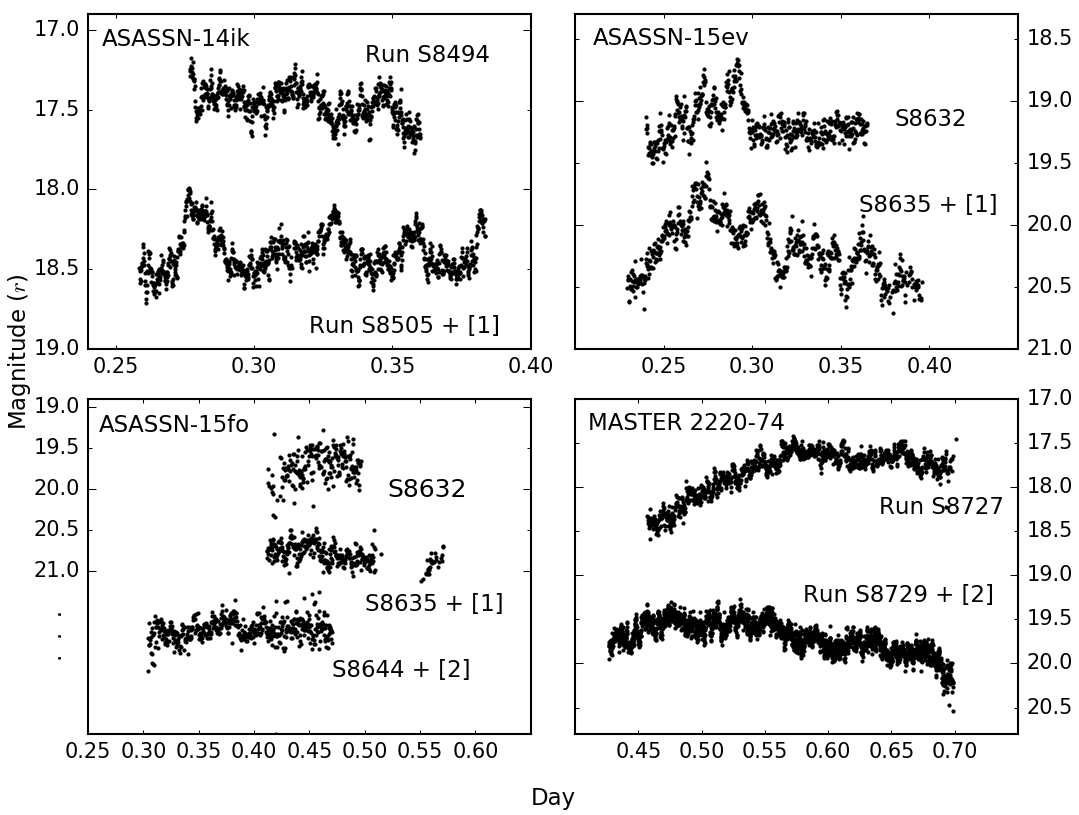}
\caption{Light curves of systems where for which periods could not be determined. Top left: Individual light curves of ASASSN-14ik (run S8494 and S8505), plotted on the same scale for comparison. Flickering with an amplitude on the order of 0.1 mag is seen both before and after the outburst, while flaring with an amplitude on the order of 0.6 mag is only seen after the outburst. Top right: Individual light curves of the long two runs on ASASSN-15ev. The light curve for run S8632 is displayed at the correct brightness; the vertical offset for S8635 is given in brackets. Bottom left: Individual light curves of ASASSN-15fo. The light curve for run S8630 is displayed at the correct brightness; the vertical offset for each light curve thereafter, is given in brackets. Bottom right: Individual light curves of the two longest runs of MASTER 2220--74. These runs show evidence of an orbital period longer than the individual runs presented in this paper.}
\label{fig:no-period}
\end{figure*}

This section contains the details of individual systems in for which \textcolor{black}{no persistent periods} were found. Some of the individual light curves for these systems are shown in Figure \ref{fig:no-period}. Our observations of the 2014 Nov outburst of ASASSN-14ik are displayed in Figure \ref{fig:long}, to show the overall shape of the observed outburst.

\subsection{ASASSN-14ik}
CRTS data of ASASSN-14ik show evidence of previous outbursts, possible superoutbursts, and high variability during quiescence. ASASSN-14ik was announced as a CV candidate by the ASAS-SN survey on 2014 October 1, when it underwent an outburst reaching $V = 14.15$ mag \textcolor{black}{\citep{2014AAS...22323603S}}, and has shown regular outbursts, occurring approximately once a month, since its discovery. Our observations in November, 2014 saw ASASSN-14ik undergoing a normal outburst, with a 2 mag amplitude and duration of 5 days. The shape of the outburst is shown in Figure \ref{fig:long}. No DNOs or QPOs were found during the outburst. Our individual light curves of two long runs (S8494, taken just before the Nov 2014 outburst, and S8505, taken after the outburst) are shown in Figure \ref{fig:no-period}. ASASSN-14ik shows flickering with an amplitude of $\sim$ 0.1 mag, as well as large flaring with an amplitude of $\sim$ 0.5 mag which is seen most clearly after the outburst. However, no persistent period was found in the data. A longer term study while the system is in quiescence is needed to confirm the presence or absence of any persistent periods.

\subsection{ASASSN-15ev}
ASASSN-15ev was announced as a CV candidate by the ASAS-SN survey on 2015 March 16, when it went into outburst with a peak magnitude of $V = 14.71$ mag \textcolor{black}{\citep{2014AAS...22323603S}}. The individual light curves of the longest two runs (S8632 and S8635) are displayed in Figure \ref{fig:no-period}, each being vertically offset for display purposes, and shows flaring of $\sim$ 0.5 mag. Matches to GALEX \citep{2018Ap&SS.363...56B} and Swift \citep{2013yCat.9043....0E} show the presence of UV and X-ray emission from ASASSN-15ev. The observations presented in this paper were taken over a month after the outburst recorded by \cite{2016PASJ...68...65K}, once the system was in quiescence. Using the relation between superhump period and orbital period of  $P_{\mathrm{orb}} = 0.9162(\pm52)P_{\mathrm{SH}} + 5.39(\pm52)$ mins, found by \cite{2009MNRAS.397.2170G}, an estimate of the orbital period can be made using the superhump period found by \cite{2016PASJ...68...65K}. With a reported superhump period of 0.057961 d, we predict an orbital period of around 0.056847 d. No evidence of the predicted orbital period, or any other periods, was found during quiescence.

\subsection{ASASSN-15fo}
ASASSN-15fo was announced as a CV candidate by the ASAS-SN survey on 2015 March 20, when it went into outburst with a peak magnitude of $V = 14.57$ mag \textcolor{black}{\citep{2014AAS...22323603S}}. The individual light curves are displayed in Figure \ref{fig:no-period}, each being vertically offset for display purposes. The observations presented in this paper were taken a month after the outburst recorded by \cite{2016PASJ...68...65K}, once the system was in quiescence. Using the relation between superhump period and orbital period found by \cite{2009MNRAS.397.2170G}, we predict an orbital period of around 0.058991 days from the superhump period of 0.060301 d report by \cite{2016PASJ...68...65K}. No evidence of the predicted orbital period, or any other periods, was found during quiescence.

\subsection{MASTER 2220--74 (MASTER OT J222049.51--740240.9)}
MASTER 2220--74 was discovered by MASTER-SAAO when it went into outburst with an amplitude of more than 3.5 mag \textcolor{black}{\citep{2015ATel.7521....1S}}. Archival data from CRTS show evidence of variability. Our individual light curves of the two longest runs are shown in Figure \ref{fig:no-period}. MASTER 2220--74 shows flickering with an amplitude on the order of 0.4 mag and a possible suggestion of a very long orbital period of over 9 hrs.

\section{Discussion and Conclusions} \label{sec:sec7}
We observed \textcolor{black}{25} CVs with the aim of classifying them, determining their orbital periods, searching for sub-orbital periodicities and highlighting interesting targets for possible in-depth studies. This sample consists of 15 CVs detected by ASAS-SN, 2 by MASTER and 8 by CRTS. A summary of the results are shown in Table \ref{tab:sumtable}.\\

\textcolor{black}{Eleven} of the systems (ASASSN-14hq, ASASSN-14ka, ASASSN-15fm, ASASSN-15pb, ASASSN-15pw, CSS 0524+00, MASTER 0014--56, MLS 0720+17, SSS 0522--35, SSS 0945--19, SSS 1340--35) were found to be eclipsing, \textcolor{black}{most} with eclipse depths $\geq$ 1 mag. Systems with clearly defined eclipse components (bright spot, accretion disc, white dwarf and donor star) can be modelled to accurately determine the systems parameters, such as masses and radii of the stellar components. This information contributes towards completing the sample distribution of CV parameters (such as orbital period distribution or white dwarf mass distribution) and plays an important role in understanding their evolution \citep{2017MNRAS.465.4968H}.
ASASSN-14ik and ASASSN-14ka have outburst periods of $\sim$ 1 month, while ASASSN-14hv has outbursts approximately every 2 months, along with superoutbursts,.\\

The light curve and periodogram of ASASSN-15fm indicate that this system is likely an IP, but further observations are needed to confirm the spin period of the white dwarf.
With an orbital period within the period range of known SW Sex CVs, and a radial velocity phase shift of 0.175$\pm$ 0.031 cycles with respect to the orbital phase, it is likely that MLS 0720+17 is a SW Sex-type CV. Although S-waves are not visible in our spectra, this is most likely due to low signal-to-noise. Higher S/N observations are also needed to confirm the presence of the phase 0.5 absorption feature seen in most SW Sex CVs.
ASASSN-15kw was found to have different photometric and spectroscopic periods, similar to GW Lib and 3 other dwarf novae. The cause of this phenomenon is still unknown, but the addition of ASASSN-15kw has increased the number of the systems showing this phenomenon to 5.\\

Out of the 5 systems that were previously observed in outburst, we were able to confirm the superhump period for 2 of the systems and obtained an orbital period for a third system while in quiescence (ASASSN-14eq).
With a superoutburst amplitude of more than 6 mag, and a superoutburst duration $\geq$ 46 days (assuming a constant decline of 0.13 mag per day and an upper limit of 21 mag for quiescence), ASASSN-17fz is classified as a WZ Sge-type star (see \citealt{2015PASJ...67..108K} for a review). Superoutburst are rare in WZ Sge-type stars, with recurrence times on the order of 1000's of days.
Although a period could not be determined for MASTER 2220--74, a suggestion of a very long orbital period, $\geq$ 9 hrs, is seen in the light curves. Long-term observations are needed to determine the orbital period.\\

\textcolor{black}{In the final paper of this series on high speed photometry of faint cataclysmic variables we reflect briefly on nearly two decades of this survey. In the nine survey papers we have presented high speed photometry of 124 CVs, with an initial focus on faint southern nova remnants and a later focus on faint CVs discovered in optical transient surveys, probing the underlying orbital period distribution of CVs. In the last three papers alone (\citealt{2012MNRAS.421.2414W}; \citealt{2014MNRAS.437..510C}, and this paper) we presented photometry of 65 CVs resulting in 43 new photometric periods. Highlights from the survey include the discovery of a fair number of new AM CVn systems including the 10-min binary ES Ceti \citep{2002PASP..114..129W}, and new insights in the nature of dwarf nova oscillations and quasi-periodic oscillations in CVs \citep{2004PASP..116..115W}. }

\begin{table*}
\caption{Summary of results.}
\label{tab:sumtable}
\begin{tabular}{lccccl}
\hline
\\
Object & Type & $P_{\mathrm{orb}}$ & $P_{\mathrm{SH}}$ & $r$ & Remarks\\
\\
\hline
\\
\textbf{ASASSN-14eq} & SU & 0.0813$(\pm3)$ & 0.079467$^c$ & 15.6 - 18.5$^a$ & Negative superhump excess\\
\textbf{ASASSN-14hq} & DN &  0.074327$(\pm9)$ & - & 18.8 - 21.7$^a$ & Eclipsing\\
\textbf{ASASSN-14hv} & SU & 0.079095$(\pm8)$ & 0.082$(\pm2)$ & 17.7 - 18.5$^a$ & Outburst $\sim$ once every two months\\
\textbf{ASASSN-14ik} & DN & - & - & 17.0 - 18.1$^a$ & Outburst $\sim$ once a month\\
\textbf{ASASSN-14ka} & DN & 0.17716$(\pm1)$ & - & 16.3 - 17.8$^a$ & Eclipsing; outburst $\sim$ once a month\\
\textbf{ASASSN-15ev} & SU & - & 0.057961$^d$ & 18.0 - 20.3$^a$ &\\
\textbf{ASASSN-15fm} & IP & 0.286$(\pm1)$ & - & 19.4 - 20.4$^a$ & Eclipsing; probable intermediate polar\\
\textbf{ASASSN-15fo} & SU & - & 0.060301$^d$ & 18.7 - 23.0$^a$ &\\
\textbf{ASASSN-15hm} & SU & - & 0.0562$(\pm1)$, 0.056219$^d$ & 14.7$^b$ & \\
\textbf{ASASSN-15hn} & SU & - & 0.06$(\pm1)$ & 14.3$^b$ & \\
\textbf{ASASSN-15kw} & DN & 0.05924$(\pm10)$ & - & 16.8 - 17.8$^a$ & Longer photometric period present\\
& & & & & alongside spectroscopic orbital period.\\
\textbf{ASASSN-15ls} & DN & 0.051$(\pm8)$ & - & 16.6 - 17.7$^a$ &\\
\textbf{ASASSN-15pb} & DN & 0.09329$(\pm2)$ & - & 18.2 - 21.1$^a$ & Eclipsing\\
\textbf{ASASSN-15pw} & DN & 0.1834$(\pm3$) & - & 16.8 - 20.3$^a$ & Eclipsing\\
\textbf{ASASSN-17fz} & WZ Sge & - & 0.05757$(\pm5)$ & 21$^a$, 14.7$^b$ & Superoutburst of $\geq$ 6 mag, slow decline\\
\textbf{CSS 0353-03} & DN &  0.0582$(\pm1)$ & - & 17.7 - 18.9$^a$ & \\
\textbf{CSS 0524+00} & DN &  0.1747$(\pm3)$ & - & 17.4 - 18.9$^a$ & Eclipsing\\
\textbf{CSS 2144+22} & SU & 0.154$(\pm1)$ & - & 16.4 - 17.3$^a$ & \\
\textbf{MASTER 0014-56} & DN & 0.0715296$(\pm6)$ & - & 19.1 - 22.9$^a$ & Eclipsing\\
\textbf{MASTER 2220-74} & DN & 0.39277$(\pm6)$ & - & 16.9 - 18.7$^a$ &\\
\textbf{MLS 0720+17} & SW Sex & 0.150409$(\pm7$) & - & 17.9 - 21.1$^a$ & Eclipsing\\
\textbf{SSS 0522-35} & DN & 0.0622$(\pm5)$ & - & 17.9 - 20.7$^a$ & \\
\textbf{SSS 0553-52} & DN & 0.0718$(\pm2)$ & - & 16.9 - 17.4$^a$ & \\
\textbf{SSS 0945-19} & SU & 0.0657693$(\pm3$) & - & 16.43 - 19.41$^a$ & Eclipsing\\
\textbf{SSS 1340-35} & DN & 0. 0598$(\pm1)$ & - & 18.42 - 19.57$^a$ & Eclipsing\\
\\
\hline
\multicolumn{6}{l}{Notes: DN: dwarf nova; SU: SU Ursae Majoris; IP: intermediate polar;
$^a$: quiescent magnitude range; $^b$:peak outburst magnitude;}\\
\multicolumn{6}{l}{$^c$: period determined by \cite{2015PASJ...67..105K};
$^d$: period determined by \cite{2016PASJ...68...65K}; $r$: $r$ magnitude of the system in quiescence.}\\
\multicolumn{6}{l}{This magnitude is an estimate and is accurate to 0.1 mag.}
\end{tabular}
\end{table*}

\section*{Acknowledgements}
We thank the anonymous referee for the suggestions and comments which helped to improve this paper.
KP acknowledges funding by the National Astrophysics and Space Science Programme (NASSP), the National Research Foundation of South Africa (NRF) through a South African Radio Astronomy Observatory (SARAO) bursary, and University of Cape Town (UCT). PW and BW acknowledge support from the NRF and UCT. JRT and CG acknowledge support from NSF grant AST1008217; and would like to thank Dartmouth undergraduates Mackenzie Carlson, Edrei Chua, Robert Cueva, Natalia Drozdoff, John French, Emma Garcia, Zoe Guttendorf, Rachel McKee, Krystyna Miles, Jack Neustadt, Sam Rosen, Marie Schwalbe, and Nick Scwhartz, all of whom helped acquire the SAAO data for MLS 0720+17 as part of a foreign study program, Dartmouth graduate students Erek Alper and Mackenzie Jones who contributed as teaching assistants to the success of the observing; as well as Prof. Brian Chaboyer for his contributions to the foreign study program. This research uses observations made at the SAAO and MDM Observatory. JRT thanks David Buckley for help with the SAAO proposal process. We acknowledge additional observations taken by D. L. Holdsworth. We acknowledge the use of the ASAS-SN, MASTER and CRTS databases. We acknowledge ESA Gaia, DPAC and the Photometric Science Alerts Team (http://gaia.ac.uk/selected-gaia-science-alerts). The national facility capability for SkyMapper has been funded through ARC LIEF grant LE130100104 from the Australian Research Council, awarded to the University of Sydney, the Australian National University, Swinburne University of Technology, the University of Queensland, the University of Western Australia, the University of Melbourne, Curtin University of Technology, Monash University and the Australian Astronomical Observatory. SkyMapper is owned and operated by The Australian National University's Research School of Astronomy and Astrophysics. The survey data were processed and provided by the SkyMapper Team at ANU. The SkyMapper node of the All-Sky Virtual Observatory (ASVO) is hosted at the National Computational Infrastructure (NCI). Development and support the SkyMapper node of the ASVO has been funded in part by Astronomy Australia Limited (AAL) and the Australian Government through the Commonwealth's Education Investment Fund (EIF) and National Collaborative Research Infrastructure Strategy (NCRIS), particularly the National eResearch Collaboration Tools and Resources (NeCTAR) and the Australian National Data Service Projects (ANDS). The Pan-STARRS1 Surveys (PS1) and the PS1 public science archive have been made possible through contributions by the Institute for Astronomy, the University of Hawaii, the Pan-STARRS Project Office, the Max-Planck Society and its participating institutes, the Max Planck Institute for Astronomy, Heidelberg and the Max Planck Institute for Extraterrestrial Physics, Garching, The Johns Hopkins University, Durham University, the University of Edinburgh, the Queen's University Belfast, the Harvard-Smithsonian Center for Astrophysics, the Las Cumbres Observatory Global Telescope Network Incorporated, the National Central University of Taiwan, the Space Telescope Science Institute, the National Aeronautics and Space Administration under Grant No. NNX08AR22G issued through the Planetary Science Division of the NASA Science Mission Directorate, the National Science Foundation Grant No. AST-1238877, the University of Maryland, Eotvos Lorand University (ELTE), the Los Alamos National Laboratory, and the Gordon and Betty Moore Foundation. Some of the data presented in this paper were obtained from the Mikulski Archive for Space Telescopes (MAST). STScI is operated by the Association of Universities for Research in Astronomy, Inc., under NASA contract NAS5-26555.

\bibliography{MN-18-2647-MJ.R1}

\bsp
\label{lastpage}
\end{document}